\documentclass[11pt]{article}
\usepackage[final]{acl}
\usepackage{times}
\usepackage{latexsym}
\usepackage[T1]{fontenc}
\usepackage[utf8]{inputenc}
\usepackage{microtype}
\usepackage{inconsolata}
\usepackage{amsmath,amssymb}
\usepackage{graphicx}
\usepackage{booktabs}
\usepackage{multirow}
\usepackage{xspace}
\usepackage{enumitem}
\usepackage{tabularx}
\usepackage{caption}
\usepackage{colortbl}
\usepackage{array}
\usepackage{subcaption}
\usepackage{algorithm}
\usepackage{algorithmic}
\usepackage{adjustbox}
\newcommand{\ours}{\textsc{LLM-Codec}\xspace}

\title{LLM-Codec: Neural Audio Codec Meets Language Model Objectives}

\author{Ho-Lam Chung$^{\ast, \ddag}$\quad Yiming Chen$^{\ddag}$\thanks{Corresponding author.} \quad Hung-yi Lee$^{\dag}$ \\
$^\ast$Graduate Institute of Communication Engineering, National Taiwan University \\
$^\dag$NTU Artificial Intelligence Center of Research Excellence (NTU AI-CoRE) \\
$^\ddag$ASUS Intelligent Cloud Services \\
}

\begin{document}
\maketitle
\begin{abstract}
Neural audio codecs are widely used as tokenizers for spoken language models, but they are optimized for waveform reconstruction rather than autoregressive prediction.
This mismatch injects acoustically driven uncertainty into the discrete token space and increases language-model perplexity.
We propose \ours, which augments codec training with language-model-facing objectives while keeping both codec and LLM architectures unchanged.
\ours introduces (i) future token prediction with Medusa-style multi-step heads to encourage multi-step predictability, and (ii) semantic alignment that matches audio and text representations via a memory-bank contrastive loss.
A differentiable Gumbel bridge enables end-to-end gradients from these objectives to the codec encoder.
On SALMon speech coherence, token LMs trained on \ours reach 61.6\% accuracy (+12.1 points over AUV) while reducing perplexity 35$\times$.
On Codec-SUPERB-tiny, \ours improves speech Mel distance by 5.0\% over AUV while simultaneously achieving the learnability gains, demonstrating that reconstruction fidelity and token predictability can be improved together.
\footnote{Code \& Model will be released at \url{https://github.com/voidful/llm-codec}}
\end{abstract}
\section{Introduction}
Following the success of large language models (LLMs),
spoken language models (SLMs) have emerged as a promising paradigm for speech generation.
SLMs represent speech as discrete token sequences and model them with autoregressive LLM backbones.
Most SLMs adopt neural audio codecs, such as EnCodec~\cite{defossez2022high} and SoundStream~\cite{zeghidour2021soundstream}, which provide a bidirectional mapping between waveforms and discrete speech tokens.
This design provides a unified interface for modeling speech and text within the same vocabulary space.

However, a fundamental tension exists between how codecs and LLMs are trained.
Codecs are optimized for \textit{reconstruction}, i.e., minimizing distortion between the waveform $x$ and its reconstruction $\hat{x}$ (e.g., $\|x-\hat{x}\|$), whereas LLMs are optimized for \textit{prediction}, i.e., maximizing next-token likelihood.
These objectives favor different representations.
To achieve high-fidelity reconstruction, a codec must preserve fine-grained acoustic factors such as pitch micro-variations, phase, breathing, and background conditions.
Many of these factors are weakly tied to linguistic content.
In a discrete token space, these factors behave like stochastic variations.
This stochasticity increases token entropy and makes the resulting sequences harder for an LLM to model.
In generation, the mismatch can manifest as repetition, semantic drift, and hallucinated content.

This paper asks a simple question. Can we retrain an existing codec encoder so that its discrete tokens remain reconstructable, but become predictable under autoregressive language modeling?

These observations suggest a simple principle: if an LLM must predict speech tokens, then the codec should be trained to emit tokens that are predictable under language modeling while preserving linguistic content.
Motivated by this, we propose \ours, which augments codec training with two LLM-facing regularizers.

First, we introduce \textbf{Future Token Prediction (FTP)}, which attaches $K$ auxiliary heads to predict multiple future tokens, implemented with a Medusa-style~\cite{cai2024medusa} design and inverse-distance weighting.
Unlike standard next-token prediction, which models only one-step dependencies, FTP encourages longer-range structure by capturing linguistic units (e.g., phonemes and words) that often span multiple tokens.

Second, \textbf{Semantic Alignment (SA)} addresses the risk of producing predictable but semantically arbitrary codes by aligning speech and text representations within the LLM using layer-wise cosine alignment and a memory-bank contrastive objective.

Finally, to enable end-to-end optimization through vector quantization, we further introduce a differentiable Gumbel-Softmax bridge that preserves discrete tokens in the forward pass while providing smooth gradients in the backward pass.

We conduct two complementary evaluations on speech coherence and audio reconstruction benchmarks.
On SALMon speech coherence benchmark, \ours produces tokens that are more amenable to language modeling: LMs trained on \ours tokens reach 61.6\% overall accuracy, substantially exceeding all baselines (48--50\%).
Moreover, \ours reduces token-level perplexity by 35$\times$ compared to the base codec, confirming that objectives, not model capacity, determine token predictability.
On Codec-SUPERB-tiny~\cite{wu2024codecsuperb}, which measures codec reconstruction quality, \ours improves speech Mel distance by 5.0\% over AUV while maintaining competitive perceptual quality.
This result demonstrates that improved token learnability does not require sacrificing signal fidelity.
Finally, \ours is simple to adopt, as it modifies only the training objectives without changing model architectures.

Overall, our major contributions include:
\begin{itemize}
\item \textbf{Objective mismatch.} We formalize the mismatch between reconstruction-trained codecs and prediction-trained LMs, and we show that it inflates audio-token uncertainty and LM perplexity.
\item \textbf{\ours Training Framework.} We propose \ours training framework, which utilizes FTP and SA as complementary regularizers to address predictability and semantics respectively. We use a hard Gumbel-Softmax bridge to backpropagate through quantization while keeping discrete tokens in the forward pass.
\item \textbf{Comprehensive Evaluation.} On SALMon, \ours improves speech-coherence accuracy to 61.6\% (+12.1 over AUV). Token-level perplexity drops 35$\times$. On Codec-SUPERB-tiny, speech Mel distance improves by 5.0\%, and ablation confirms that reconstruction and learnability gains are additive.
\end{itemize}

\section{Related Work}

\paragraph{Neural audio codecs.}
SoundStream~\cite{zeghidour2021soundstream} and EnCodec~\cite{defossez2022high} established neural audio compression with vector quantization and adversarial training.
BigCodec~\cite{xin2024bigcodec} explores a large single codebook.
WavTokenizer~\cite{ji2024wavtokenizer} improves codebook utilization and targets better language modeling compatibility.
These methods optimize reconstruction quality.
\ours modifies the training objective to incorporate language-model-facing signals.

\paragraph{Semantic speech tokens and self-supervised learning.}
Self-supervised models such as wav2vec 2.0~\cite{baevski2020wav2vec} and HuBERT~\cite{hsu2021hubert} learn discrete or pseudo-discrete units that correlate with phonetic content.
SpeechTokenizer~\cite{zhang2024speechtokenizer} and related lines separate semantic and acoustic factors for speech LLMs.
These approaches often introduce additional encoders or decoders.
\ours keeps the codec architecture unchanged and instead reshapes the codec via LLM-aligned losses.

\paragraph{Multi-step prediction.}
Medusa~\cite{cai2024medusa} introduces multiple decoding heads to predict several future tokens for faster LLM inference.
FlowSLM~\cite{flowslm2025} argues that spoken language modeling benefits from constraints beyond one-step prediction.
\ours adapts the Medusa idea as a training-time regularizer to encourage multi-step predictability in codec tokens.

\paragraph{Audio-text alignment.}
CLAP~\cite{elizalde2023clap} aligns audio and text for retrieval in a shared embedding space.
SpeechGPT~\cite{zhang2023speechgpt} projects speech features into LLM input space for dialogue.
\ours aligns audio and text representations inside the hidden layers of a frozen LLM.
This choice directly targets generation and token predictability.

\paragraph{Token consistency and factorized tokenizers.}
Recent work studies token instability under acoustic perturbations.
\citet{liu2025analyzing} identify the Discrete Representation Inconsistency (DRI) phenomenon, attributing it to encoder receptive field leakage across contexts, and propose consistency regularizers to mitigate it.
\citet{wu2025towards} propose codec-LM co-design techniques including a framewise encoder and codebook-level dropout to improve end-to-end TTS performance.
Other lines decouple semantic and acoustic factors, or use spectral quantization to obtain more predictable token sequences by construction.
\ours is complementary: it diagnoses a different root cause, the objective mismatch between reconstruction and prediction, and addresses it through LLM-facing training objectives without changing model architectures.

\section{Preliminaries}
\subsection{Neural Audio Codecs}
A neural audio codec compresses audio into discrete tokens.
Let $\mathcal{E}$ be the encoder, $\mathcal{Q}$ the quantizer, and $\mathcal{D}$ the decoder.
Given waveform $x \in \mathbb{R}^T$, the codec produces:
\begin{equation}
z = \mathcal{E}(x), \quad c = \mathcal{Q}(z), \quad \hat{x} = \mathcal{D}(c).
\end{equation}
The training objective is reconstruction:
\begin{equation}
\mathcal{L}_{\text{codec}} = \mathcal{L}_{\text{recon}}(x, \hat{x}) + \mathcal{L}_{\text{VQ}}(z),
\end{equation}
where $\mathcal{L}_{\text{recon}}$ includes time-domain L1, mel-spectrogram L1, and adversarial losses.
$\mathcal{L}_{\text{VQ}}$ is the commitment loss for the quantizer.
Modern codecs like EnCodec~\cite{defossez2022high} use Residual Vector Quantization (RVQ) with multiple codebooks.
This achieves high reconstruction quality at low bitrates (e.g., 6 kbps).

\subsection{Spoken Language Models}
A spoken language model treats codec tokens as a language and models their distribution autoregressively:
\begin{equation}
p(c_1, \ldots, c_T) = \prod_{t=1}^{T} p(c_t | c_{<t}).
\end{equation}
This formulation allows SLMs to leverage powerful transformer architectures developed for text LLMs.
Systems like VALL-E~\cite{wang2023neural}, AudioLM~\cite{borsos2023audiolm}, and SpeechGPT~\cite{zhang2023speechgpt} have demonstrated impressive speech generation capabilities.

\subsection{The Objective Mismatch Problem}
The codec and the LLM are optimized for fundamentally different goals.

\textbf{Codec goal}: preserve all information needed for reconstruction, including linguistic content, speaker identity, prosody, and fine-grained acoustic details.

\textbf{LLM goal}: model sequential dependencies, which favors token sequences that exhibit predictable patterns given context.

These goals can conflict: to faithfully reconstruct acoustic nuances, a codec may assign different tokens to acoustically distinct but linguistically equivalent realizations.
For example, the same word ``hello'' can map to different token sequences under changes in pitch, speaking rate, or background noise.
From the LLM's perspective, such variation appears as noise and reduces token predictability.

\paragraph{A concrete example.}
Consider the word ``hello'' spoken twice by the same speaker.
The two utterances are linguistically identical but acoustically slightly different (different pitch contour, duration, breath).
A reconstruction-oriented codec might encode them as:
\begin{align*}
\text{``hello''}_1 &\to [1042, 3891, 2847, 1923] \\
\text{``hello''}_2 &\to [1042, 3892, 2851, 1919]
\end{align*}
The tokens are similar but not identical.
This creates two problems for the LLM.
First, the model must learn multiple token patterns for the same word, increasing the effective vocabulary complexity.
Second, during generation, even if the model correctly predicts ``hello,'' it must choose among acoustically valid variants, introducing unnecessary degrees of freedom.
These variations accumulate: a sentence has exponentially many valid token sequences, making both learning and generation harder.

We note that token variability under acoustic perturbation has been independently studied by \citet{liu2025analyzing}, who attribute it to context leakage in encoder receptive fields.
Our analysis differs in both the input condition (natural acoustic variation vs.\ contextual truncation) and the diagnosed root cause (reconstruction objective mismatch vs.\ receptive field leakage).
Both perspectives are complementary and identify the same symptom from different angles.

\subsection{Desiderata for LLM-Friendly Tokens}
Based on this analysis, we identify two properties that LLM-friendly tokens should have:

\paragraph{Property 1: Multi-step predictability.}
Given context, not just the next token but several future tokens should be predictable.
Linguistic units (phonemes, words, phrases) span multiple tokens.
If a word starts, the LLM should be able to predict how it ends.

\paragraph{Property 2: Semantic consistency.}
Tokens representing the same linguistic content should produce similar LLM representations, regardless of acoustic variations.
``Hello'' should look like ``hello'' inside the LLM, whether spoken loudly or softly.

Standard codecs satisfy neither property.
They optimize reconstruction, not predictability or semantic consistency.

\section{LLM-Codec}

\begin{figure*}[t]
\centering
\includegraphics[width=1\linewidth]{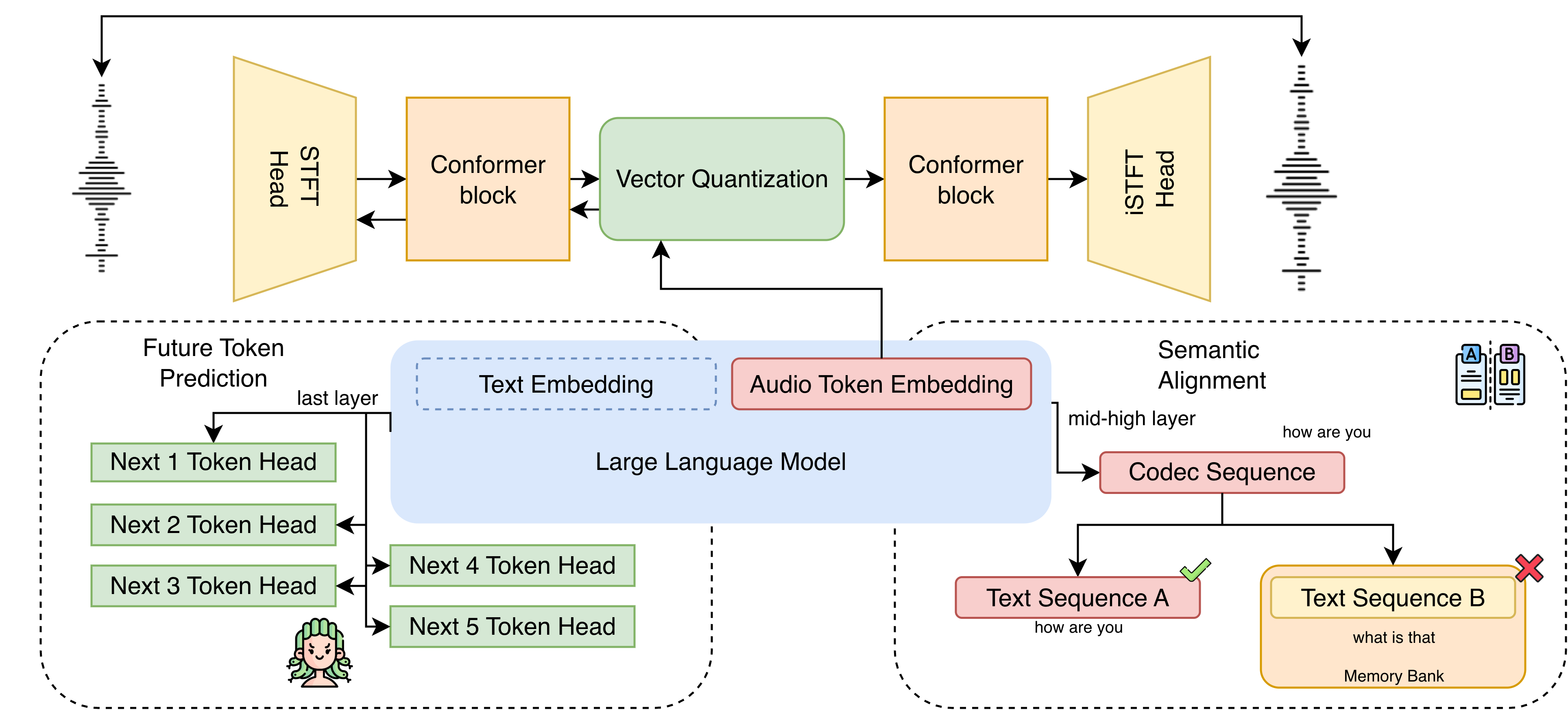}
\caption{\textbf{Overview of \ours.} 
Audio is encoded by the codec and passed through a Gumbel bridge to obtain differentiable embeddings.
A single LLM forward pass produces hidden states for both FTP (using $K$ Medusa heads) and SA (aligning with text representations).
Gradients flow back through the bridge to update the codec encoder.}
\label{fig:overview}
\end{figure*}

We propose to train the codec with LLM-facing objectives.
Figure~\ref{fig:overview} shows the overall architecture.
The key components are: (1) future token prediction heads, (2) semantic alignment losses, and (3) a differentiable Gumbel bridge.

\subsection{Future Token Prediction (FTP)}

\paragraph{Why multi-step prediction?}
Standard next-token prediction optimizes one-step dependencies.
But linguistic structure spans multiple tokens.
A phoneme at 50 Hz might be 2--4 tokens.
A word might be 5--15 tokens.
We want the codec to produce tokens where seeing the beginning of a word helps predict its ending.

\paragraph{Medusa-style heads.}
We add $K$ prediction heads, where head $M_k$ predicts the token at offset $k$.
Each head is a linear projection:
\begin{equation}
M_k: \mathbb{R}^H \to \mathbb{R}^V, \quad M_k(h) = h W_k.
\end{equation}

\paragraph{Initialization from LLM head.}
Let $W_{\text{LM}}$ denote the frozen LLM output projection.
Let $\mathcal{V}_{\text{audio}}$ be the index set of audio tokens in the extended vocabulary.
We initialize each Medusa head by copying the audio-token sub-matrix from $W_{\text{LM}}$:
\begin{equation}
W_k \leftarrow \mathrm{SelectAudio}(W_{\text{LM}}, \mathcal{V}_{\text{audio}}).
\end{equation}
This initialization leverages the pretrained output geometry and stabilizes early training.

\paragraph{Inverse-distance weighting.}
Near-future predictability should dominate the signal, but farther horizons are still informative.
We use inverse-distance weights and normalize them to sum to one:
\begin{equation}
\small
w_k = \frac{1/k}{\sum_{j=1}^{K} 1/j}.
\end{equation}
For $K=5$, this yields $w \approx [0.44, 0.22, 0.15, 0.11, 0.09]$.

\paragraph{Loss formulation.}
Let $T$ be the audio-token sequence length.
We define the FTP loss as a weighted multi-step cross entropy:
\begin{equation}
\small
\mathcal{L}_{\text{FTP}} =
\frac{1}{T-K}\sum_{t=1}^{T-K}\sum_{k=1}^{K} w_k \cdot \text{CE}(M_k(h_t), c_{t+k}),
\end{equation}
where $h_t$ is the LLM hidden state at position $t$ and $c_{t+k}$ is the target token at offset $k$.

\paragraph{Gradient flow.}
Gradients from FTP flow through the LLM's hidden states, the input embeddings, the Gumbel bridge (which will be further introduced in Section~\ref{subsec:gumbel}), and finally to the codec encoder.
This end-to-end gradient flow is what allows FTP to shape the codec's behavior.

\subsection{Semantic Alignment (SA)}

\paragraph{Why semantic alignment?}
FTP encourages predictability, but predictability alone is not enough.
A codec could learn to produce predictable but semantically meaningless tokens.
We need to ensure tokens preserve linguistic content.

\paragraph{Core idea.}
If audio and text represent the same content, they should produce similar representations inside the LLM.
We enforce this by aligning hidden states from the audio branch with hidden states from the text branch.

\paragraph{Layer selection.}
Not all layers are equally suitable for alignment.
Lower layers capture modality-specific surface features.
Upper layers capture abstract semantics.
Let $L$ denote the total number of layers in the LLM.
We align middle-to-high layers: $l \in [L/3, 0.8L]$.
For a 32-layer LLM, this corresponds to layers 10--25.

\paragraph{Representation extraction.}
We align sequence-level states because speech and text have different token rates and lengths.
For a causal transformer, the last hidden state summarizes the full prefix.
For each selected layer $l$, we use last-position pooling:
\begin{itemize}[leftmargin=1.2em, itemsep=0em, topsep=0.1em]
\item Audio: $h_{\text{audio}}^{(l)}$ is the hidden state at the last audio token.
\item Text: $h_{\text{text}}^{(l)}$ is the hidden state at the last non-padding text token.
\end{itemize}
We compute the text branch with \texttt{no\_grad} and detach $h_{\text{text}}^{(l)}$.
This prevents the audio pathway from drifting the semantic geometry of text representations and stabilizes the alignment objective.

\paragraph{Cosine alignment loss.}
We minimize cosine distance across selected layers:
\begin{equation}
\small
\mathcal{L}_{\text{cos}} = \frac{1}{|\mathcal{L}|} \sum_{l \in \mathcal{L}} \left(1 - \cos(\hat{h}_{\text{audio}}^{(l)}, \hat{h}_{\text{text}}^{(l)}) \right),
\end{equation}
where $\hat{h}$ denotes $\ell_2$-normalized vectors.

\paragraph{Contrastive loss with memory bank.}
Cosine loss alone can cause representation collapse.
We add contrastive learning to maintain discriminability.
We maintain a memory bank $Q$ of recent text representations (FIFO queue, size 512).
For each audio representation, the paired text is positive, and bank entries are negatives:
\begin{equation}
\scalebox{0.7}{$\displaystyle
\mathcal{L}_{\text{ctr}} = -\log \frac{\exp(\alpha \cdot \text{sim}(h_a, h_t^+))}{\exp(\alpha \cdot \text{sim}(h_a, h_t^+)) + \sum_{q \in Q} \exp(\alpha \cdot \text{sim}(h_a, q))}
$}
\end{equation}
where $\alpha = 5.0$ is the logit scale and we apply label smoothing with $\epsilon = 0.1$.

\subsection{Differentiable Gumbel Bridge} \label{subsec:gumbel}
Vector quantization uses argmax, which has zero gradient almost everywhere.
We use the Gumbel-Softmax trick~\cite{jang2017categorical} with hard sampling: discrete tokens in the forward pass, smooth gradients in the backward pass.
Let $z_t \in \mathbb{R}^C$ be the codec's continuous latent at time $t$.
A linear projection maps it to logits $\ell_t = z_t W_{\text{bridge}}$, and we apply $y_t = \text{GumbelSoftmax}(\ell_t / \tau, \text{hard=True})$.
The LLM embedding is $e_t = y_t E_{\text{audio}}$, where $E_{\text{audio}} \in \mathbb{R}^{V \times H}$ is the audio token embedding matrix.
We anneal temperature $\tau$ from 1.0 to 0.3 over 20k steps.
To prevent the bridge from diverging from the codec's quantizer, we add $\mathcal{L}_{\text{bridge}} = \text{CrossEntropy}(\ell_t, c_t)$, where $c_t$ is the codec's original token.

\subsection{Training Procedure}
We activate LLM-facing objectives only after codec reconstruction stabilizes, using a delayed ramp-up schedule to avoid injecting high-variance gradients into early training.
We stagger the two objectives so the codec first adapts to token predictability (FTP), then to semantic alignment (SA).
FTP activates at step 10k and ramps to full weight by step 12k.
SA activates at step 12k and ramps to full weight by step 14k.
This staggering avoids gradient interference between the two objectives during initial ramp-up.
The total loss combines the standard codec objective $\mathcal{L}_{\text{codec}}$, the bridge alignment $\mathcal{L}_{\text{bridge}}$, and the step-weighted LLM regularizers $\mathcal{L}_{\text{FTP}}$, $\mathcal{L}_{\text{cos}}$, and $\mathcal{L}_{\text{ctr}}$.

We optimize:
\begin{equation}
\small
\begin{split}
\mathcal{L}_{\text{total}} ={}
& \mathcal{L}_{\text{codec}} +
\lambda_{\text{bridge}}\mathcal{L}_{\text{bridge}} +
\lambda_{\text{FTP}}\mathcal{L}_{\text{FTP}} \\
& + \lambda_{\text{cos}}\mathcal{L}_{\text{cos}} +
\lambda_{\text{ctr}}\mathcal{L}_{\text{ctr}}.
\end{split}
\end{equation}
We train the codec encoder and decoder, Gumbel bridge, audio token embeddings, and Medusa heads, while freezing the LLM backbone (to preserve text capability).
We use a substantially lower learning rate for the decoder than for the auxiliary components, so that the reconstruction pathway remains stable while the encoder adapts to LLM-facing signals.
The active reconstruction loss further constrains the decoder to its pretrained operating region.
Inference cost is unchanged because auxiliary heads are discarded at test time.

\section{Experiments}

\subsection{Experiment Setup}

\paragraph{Codec training.}
We train on LibriSpeech train-clean-100 with paired transcripts~\cite{7178964}.
We fine-tune the AUV~\cite{chen2025auv} codec (both encoder and decoder) at 50 Hz (vocabulary size 20{,}480).
We use Qwen3-4B-Instruct~\cite{yang2025qwen3} (32 layers, hidden size 2{,}560) as a frozen LLM backbone to isolate the effect of tokenization.
We train the codec for 25k steps with 4-second segments and an effective batch size of 10.
Appendix~\ref{app:impl} reports the full configuration and hyperparameters.

\paragraph{SLM training.}
We use the same LM training recipe across tokenizers.
We tokenize LibriSpeech train-clean-100 with each codec.
We fine-tune Qwen3-4B-Instruct~\cite{yang2025qwen3} with LoRA~\cite{hu2022lora} (rank 64, $\alpha$=128, dropout 0.05, applied to all linear layers) using next-token prediction with learning rate $10^{-4}$ and effective batch size 32.
We use a cosine learning rate scheduler with 10\% warmup.
For \ours, we start from the codec-trained checkpoint which already contains the expanded audio vocabulary and trained audio-token embeddings.
For baselines, we expand the vocabulary at the start of SLM training.
In both cases, the LLM backbone is Qwen3-4B-Instruct, and LoRA adapts all linear layers identically.
We train for 3 epochs and evaluate on SALMon~\cite{nachmani2023salmon}.

\paragraph{SLM evaluation.}
We measure token learnability by training a token-level speech LM and evaluating coherence on SALMon.
SALMon evaluates whether a model assigns higher likelihood to coherent speech than to minimally perturbed incoherent variants.
Each example contains a coherent sample and an incoherent counterpart constructed by changing one acoustic factor mid-utterance.
We select the coherent sample by comparing length-normalized negative log-likelihood under the token LM.
We report the speaker and acoustic environment categories in the main paper, since they most directly reflect token-level stability under mid-utterance perturbations.
We report the emotion categories in the appendix.

\paragraph{Reconstruction evaluation.}
We evaluate codec reconstruction on Codec-SUPERB-tiny~\cite{wu2024codecsuperb} across Speech, Music, and Environmental Audio.
We report domain-appropriate metrics to avoid misleading comparisons.
Main-text tables use Speech (Mel, STFT, PESQ, STOI), Music (Mel, STFT, PESQ, STOI, F0), and Audio (Mel, STFT, PESQ).
Appendix~\ref{app:results} provides the full metric applicability rationale.

\paragraph{Baselines.}
We compare against strong codec tokenizers that are commonly used for speech or audio language modeling.
We match the token rate to 50 tokens/s to control sequence length and modeling difficulty.
The baselines include AUV~\cite{chen2025auv} (our starting point), BigCodec~\cite{xin2024bigcodec}, UniCodec~\cite{jiang2025unicodec}, and WavTokenizer~\cite{ji2024wavtokenizer} variants with different capacities.

\subsection{Speech Language Modeling}

Reconstruction fidelity is not sufficient for spoken language modeling.
We therefore first evaluate whether \ours produces more LLM-friendly token sequences by training token-level LMs and testing speech coherence on SALMon.

\begin{table*}[t]
\centering
\small
\begin{tabular}{l|cc|cccc|c}
\toprule
& \multicolumn{2}{c|}{\textbf{Speaker}} & \multicolumn{4}{c|}{\textbf{Acoustic Environment}} & \\
\textbf{Model} & \textbf{Spkr} & \textbf{Gend} & \textbf{RIR} & \textbf{BG-Align} & \textbf{BG-Dom} & \textbf{BG-All} & \textbf{Overall} \\
\midrule
WavTok-L & 47.0 & 52.5 & 37.5 & 51.5 & 50.5 & 51.0 & 48.3 \\
BigCodec & 50.5 & 49.5 & 43.5 & 48.0 & 53.5 & 48.5 & 49.4 \\
UniCodec & 49.0 & 53.0 & 53.0 & 47.5 & 45.5 & 46.0 & 50.1 \\
AUV & 47.5 & 52.5 & 44.0 & 45.5 & 53.5 & 49.0 & 49.4 \\
\midrule
\textbf{\ours} & \textbf{63.0} & \textbf{65.0} & \textbf{62.5} & \textbf{48.0} & \textbf{69.0} & \textbf{71.5} & \textbf{61.6} \\
\bottomrule
\end{tabular}
\caption{\textbf{SALMon speech coherence evaluation.} Accuracy (\%) on speaker and acoustic consistency after 3 epochs of LM training. Overall is the 8-category SALMon average (including sentiment, reported in Appendix). All baselines cluster around chance level (48--50\%). \ours achieves 61.6\% overall (+12.1 over AUV). Gains are largest on background consistency (BG-All: +22.5, BG-Dom: +15.5) and reverberation (RIR: +18.5).}
\label{tab:salmon}
\end{table*}

\paragraph{Main results.}
Table~\ref{tab:salmon} shows SALMon accuracy. Three findings stand out.

\textit{(1) \ours outperforms all baselines by a large margin.} \ours achieves 61.6\% overall accuracy, compared to 49.4\% for AUV (+12.1 points).
All baselines cluster near chance level (48--50\%), regardless of their reconstruction quality.

\textit{(2) Gains are broad across categories.} \ours improves over AUV on every category.
The largest gains appear on background consistency (BG-All: 71.5\% vs.\ 49.0\%, +22.5 points) and reverberation (RIR: 62.5\% vs.\ 44.0\%, +18.5 points).

\textit{(3) Reconstruction quality does not predict learnability.} UniCodec and BigCodec achieve competitive reconstruction scores (Section~\ref{sec:recon}), yet their SALMon accuracy is no better than AUV or WavTokenizer.
This confirms that the reconstruction-prediction mismatch, not codec capacity, is the bottleneck.

SALMon isolates a property that matters for token LMs, namely likelihood contrast between coherent and minimally perturbed incoherent audio.
This likelihood-based protocol directly tests whether a tokenizer produces sequences that an LM can model reliably.

\paragraph{Why do speaker-related scores improve?}
We hypothesize that SA encourages representations that are more invariant to speaker-specific nuisance factors given the same transcript.
Such invariance can make mid-utterance speaker changes appear as sharper distribution shifts, which increases likelihood contrast on SALMon.

\paragraph{Initialization matters.}
We initialize audio token embeddings from the LLM's text embedding space.
A random initialization reduces SALMon accuracy to 50.7\%, which is close to chance for a binary-choice benchmark.
This result suggests that reusing the LLM's pre-trained embedding geometry stabilizes early training for speech tokens.

\begin{figure}[t]
\centering
\includegraphics[width=0.95\linewidth]{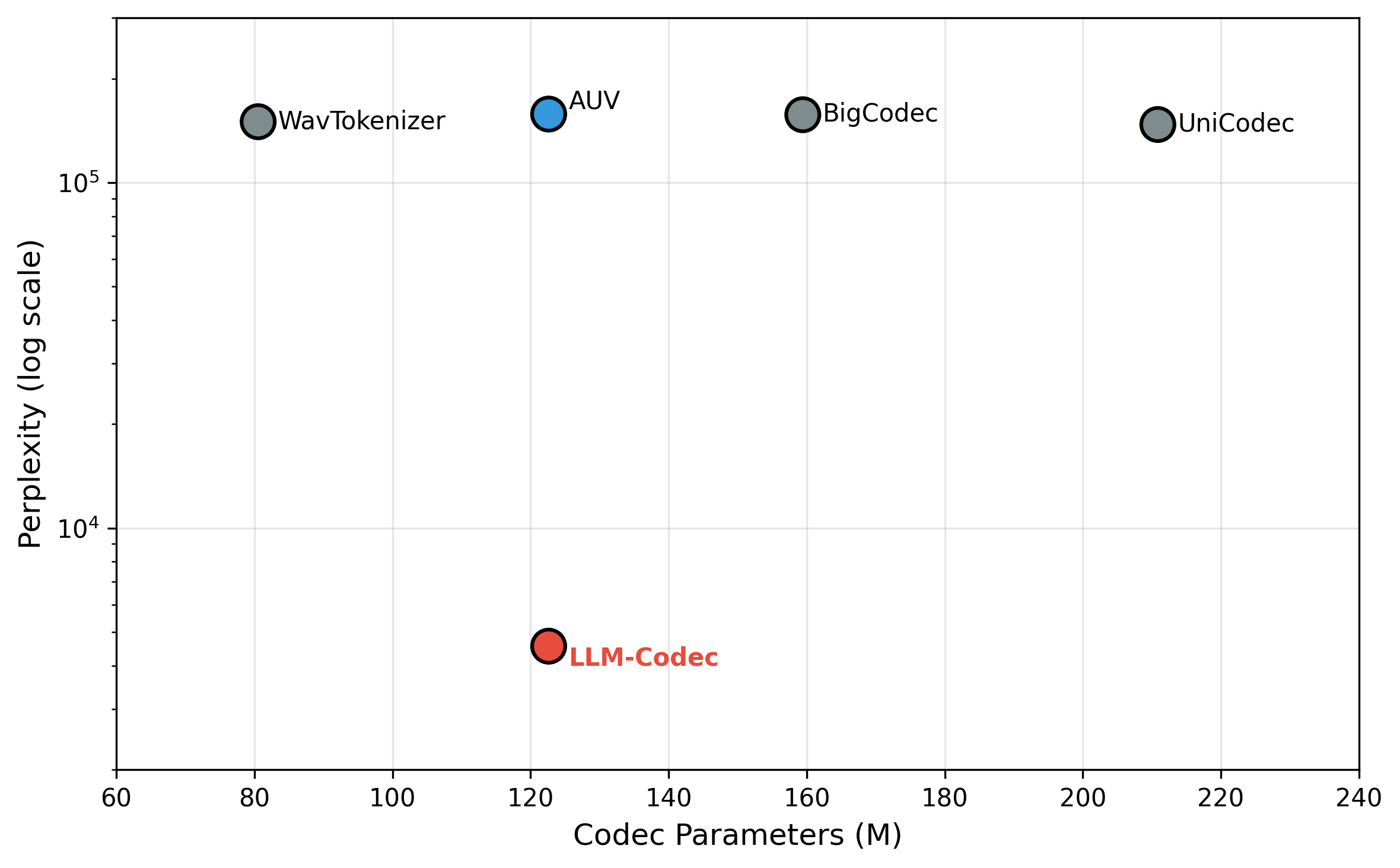}
\caption{\textbf{Perplexity is determined by training objectives, not model size.} All baselines (80M--211M parameters) achieve similar perplexity (148K--160K). \ours (122M, same as AUV) achieves 4{,}617, a 35$\times$ reduction. This confirms that the codec-LLM objective mismatch, not model capacity, is the bottleneck.}
\label{fig:ppl}
\end{figure}

\paragraph{Token predictability.}
Beyond SALMon accuracy, we directly measure token predictability via language modeling perplexity.
Figure~\ref{fig:ppl} shows validation perplexity on LibriSpeech for each codec.
All baselines exhibit perplexity in the range 148K--160K, regardless of codec parameter count (80M--211M).
In contrast, \ours achieves perplexity of 4{,}617, which is 35$\times$ lower than AUV (159{,}768) despite identical parameter counts (122.63M).
This result confirms that improvements stem from training objectives, not model capacity.
The perplexity reduction directly validates our core claim: LLM-facing objectives produce more predictable tokens.

\paragraph{Implications.}
SALMon results support our central claim that LLM-facing objectives can shape a codec tokenizer toward more learnable token sequences.
The improvement over AUV indicates that token predictability is a first-order bottleneck for spoken language modeling.

\subsection{Reconstruction Quality}
\label{sec:recon}

We next evaluate reconstruction quality to verify that LLM-facing objectives do not degrade codec fidelity.
Table~\ref{tab:main} summarizes reconstruction across domains.

\begin{table*}[t]
\begin{adjustbox}{width=1\textwidth}
\centering
\small
\begin{tabular}{l|cccc|ccccc|ccc}
\toprule
& \multicolumn{4}{c|}{\textbf{Speech}} & \multicolumn{5}{c|}{\textbf{Music}} & \multicolumn{3}{c}{\textbf{Audio}} \\
\textbf{Model} & \textbf{Mel}$\downarrow$ & \textbf{STFT}$\downarrow$ & \textbf{PESQ}$\uparrow$ & \textbf{STOI}$\uparrow$ & \textbf{Mel}$\downarrow$ & \textbf{STFT}$\downarrow$ & \textbf{PESQ}$\uparrow$ & \textbf{STOI}$\uparrow$ & \textbf{F0}$\uparrow$ & \textbf{Mel}$\downarrow$ & \textbf{STFT}$\downarrow$ & \textbf{PESQ}$\uparrow$ \\
\midrule
BigCodec & 0.810 & 1.718 & \textbf{2.208} & \textbf{0.877} & 1.522 & 3.108 & 1.922 & \underline{0.606} & 0.728 & 2.101 & 4.864 & 1.391 \\
UniCodec & 0.830 & 1.824 & 2.022 & 0.851 & \underline{1.196} & \textbf{2.441} & 1.859 & 0.588 & \underline{0.760} & \textbf{1.337} & \textbf{3.336} & 1.548 \\
WavTok-M & 0.904 & 1.846 & 1.843 & 0.823 & 1.407 & 2.683 & 1.579 & 0.528 & 0.683 & \underline{1.382} & \underline{3.367} & 1.380 \\
WavTok-L & 1.133 & 2.006 & 1.547 & 0.765 & 1.558 & 2.896 & 1.484 & 0.478 & 0.722 & 1.579 & 3.777 & 1.334 \\
WavTok-S & 1.096 & 2.174 & 1.437 & 0.761 & 1.631 & 2.976 & 1.350 & 0.475 & 0.629 & 1.487 & 3.442 & 1.273 \\
\midrule
AUV (base) & \underline{0.762} & \underline{1.648} & 2.094 & 0.850 & \underline{1.129} & 2.557 & \textbf{2.195} & \underline{0.609} & \textbf{0.785} & 1.847 & 4.407 & \textbf{1.567} \\
\textbf{\ours} & \textbf{0.724} & \textbf{1.599} & \underline{2.102} & \underline{0.859} & \textbf{1.126} & \underline{2.541} & \underline{2.175} & \textbf{0.614} & \underline{0.767} & 1.836 & 4.389 & \underline{1.562} \\
\bottomrule
\end{tabular}
\end{adjustbox}
\caption{\textbf{Reconstruction quality across domains.} Evaluated on Codec-SUPERB-tiny~\cite{wu2024codecsuperb}. All models operate at 50 tokens/s. We report domain-appropriate metrics: F0 only for Music (pitch matters for melody), STOI only for Speech/Music (intelligibility undefined for environmental sounds). \textbf{Speech}: \ours achieves best spectral fidelity (Mel 0.724, STFT 1.599), improving over AUV by 5.0\% and 3.0\% respectively. \textbf{Music}: \ours matches or slightly improves AUV. \textbf{Audio}: AUV leads on PESQ.}
\label{tab:main}
\end{table*}

\paragraph{Speech domain: \ours improves spectral fidelity.}
On speech, \ours achieves the best spectral fidelity with Mel distance 0.724 and STFT distance 1.599.
Relative to AUV, \ours improves Mel by 5.0\% and STFT by 3.0\%.
Perceptual metrics also improve slightly: PESQ from 2.094 to 2.102 and STOI from 0.850 to 0.859.
BigCodec achieves the highest perceptual scores (PESQ 2.208, STOI 0.877), but \ours closes most of the gap while leading on spectral metrics.
Combined with the 35$\times$ perplexity reduction (Section~5.2), this result demonstrates that LLM-facing training can improve both reconstruction fidelity and token learnability simultaneously.

\paragraph{Why does reconstruction improve?}
We train both the codec encoder and decoder at a low learning rate ($5 \times 10^{-6}$) with multiple reconstruction losses (mel, multi-scale mel, multi-resolution STFT, complex STFT, and GAN).
The active reconstruction losses constrain both encoder and decoder to remain within the pretrained operating region, while allowing gradual adaptation.
Our ablation (Table~\ref{tab:ablation_recon}) confirms that the reconstruction improvement comes from this training procedure, not from FTP or SA specifically.

\paragraph{Music domain: comparable to AUV.}
On music, \ours matches or slightly improves AUV.
Mel distance is marginally better (1.126 vs.\ 1.129), and STOI improves (0.614 vs.\ 0.609).
AUV retains a small advantage on PESQ (2.195 vs.\ 2.175) and F0 correlation (0.785 vs.\ 0.767).
The differences are small, confirming that our speech-centric training does not harm music reconstruction.

\paragraph{Audio (environmental) domain: comparable to AUV.}
On environmental audio, UniCodec achieves the best spectral metrics, while \ours is comparable to AUV.
\ours slightly improves Mel (1.836 vs.\ 1.847) but slightly reduces PESQ (1.562 vs.\ 1.567).

\paragraph{Domain-specific summary.}
\ours improves speech spectral fidelity by 5.0\% Mel while preserving or slightly improving reconstruction on music and environmental audio.
The improvements on speech are consistent with the training procedure, which uses speech-text supervision and GAN-based adversarial training.
The learnability gains (35$\times$ perplexity, +12.1 SALMon) are orthogonal to and come on top of these reconstruction improvements.

\subsection{Analysis}

\paragraph{Learnability versus reconstruction.}
A key finding of this work is the dissociation between reconstruction quality and token learnability.
All baselines achieve comparable reconstruction on speech (Table~\ref{tab:main}), yet their token-level LMs plateau at chance-level SALMon accuracy (Table~\ref{tab:salmon}).
\ours improves both reconstruction (5.0\% speech Mel) and learnability (+12.1 SALMon, 35$\times$ perplexity).
Importantly, our ablation shows that the reconstruction improvement comes from the training procedure (GAN, multi-scale losses), while the learnability improvement comes from FTP and SA.
These two effects are additive and orthogonal.

\paragraph{What makes tokens learnable?}
Our results suggest that learnability requires two complementary properties, both introduced by \ours.
First, tokens should be \textit{locally predictable}: given context, the next few tokens should be deterministic. FTP enforces this by penalizing unpredictable futures.
Second, tokens should be \textit{semantically grounded}: the same linguistic content should map to similar representations regardless of acoustic variation. SA enforces this by aligning audio and text branches.

\subsection{Ablation Study}

We ablate each objective on speech reconstruction to isolate the roles of FTP and SA.

\paragraph{Component ablation.}
Table~\ref{tab:ablation_recon} ablates FTP and SA on both reconstruction fidelity and token learnability.
All variants improve Mel distance by $\sim$5\% over the original AUV, but are indistinguishable from each other on reconstruction ($\leq$0.002 Mel).
This confirms that the reconstruction improvement comes from the shared training procedure (GAN, multi-scale losses), not from FTP or SA.

On learnability, all LLM-Codec variants achieve dramatically lower perplexity (4.6K vs.\ 160K) and higher SALMon accuracy (61--62\% vs.\ 49\%).
FTP-only and SA-only each achieve similar learnability to the full model, suggesting that each objective independently captures most of the benefit.

\begin{table}[t]
\centering
\small
\begin{adjustbox}{width=\columnwidth}
\begin{tabular}{lccc}
\toprule
\textbf{Variant} & \textbf{Mel}$\downarrow$ & \textbf{PPL}$\downarrow$ & \textbf{SALMon}$\uparrow$ \\
\midrule
AUV (original) & 0.762 & 159{,}768 & 49.4 \\
\midrule
FTP only & 0.725 & 4{,}631 & 61.8 \\
SA only & 0.723 & 4{,}616 & 61.3 \\
\ours (FTP + SA) & 0.724 & 4{,}617 & 61.6 \\
\bottomrule
\end{tabular}
\end{adjustbox}
\caption{\textbf{Component ablation (Speech).} All variants improve Mel $\sim$5\% over AUV (from training procedure). All variants achieve $\sim$35$\times$ lower PPL and $\sim$+12 SALMon (from LLM-facing objectives). FTP and SA each independently capture the full learnability gain.}
\label{tab:ablation_recon}
\end{table}

\paragraph{Effect of prediction horizon $K$.}
Table~\ref{tab:ablation_k} ablates the number of Medusa heads $K$ in FTP.
Reconstruction, perplexity, and SALMon are all invariant to $K$.
This indicates that even $K{=}1$ (next-token prediction only) suffices to reshape the token space for learnability, and multi-step prediction does not provide additional gains in our setting.

\begin{table}[t]
\centering
\small
\begin{adjustbox}{width=\columnwidth}
\begin{tabular}{lccc}
\toprule
\textbf{$K$} & \textbf{Mel}$\downarrow$ & \textbf{PPL}$\downarrow$ & \textbf{SALMon}$\uparrow$ \\
\midrule
AUV (original) & 0.762 & 159{,}768 & 49.4 \\
\midrule
$K=1$ & 0.725 & 4{,}561 & 61.6 \\
$K=3$ & 0.723 & 4{,}554 & 61.4 \\
$K=5$ (default) & 0.724 & 4{,}617 & 61.6 \\
$K=10$ & 0.724 & 4{,}539 & 61.4 \\
\bottomrule
\end{tabular}
\end{adjustbox}
\caption{\textbf{Effect of prediction horizon $K$ (Speech).} All metrics are invariant to $K$. Even $K{=}1$ achieves the full learnability gain, suggesting that the key factor is the presence of LLM-facing gradients, not the prediction horizon length.}
\label{tab:ablation_k}
\end{table}

\section{Discussion}

\paragraph{Why does FTP help?}
FTP directly regularizes token sequences to be predictable beyond the next step.
Our ablation confirms that FTP alone achieves most of the learnability gain (SALMon 61.8 vs.\ 49.4 for AUV).
Interestingly, even $K{=}1$ (single-step prediction) suffices, suggesting that the key mechanism is the LLM-facing gradient signal itself, not the multi-step horizon.

\paragraph{Why does SA help?}
SA injects an external semantic anchor by tying audio-induced hidden states to text-induced hidden states.
SA alone also achieves strong learnability (SALMon 61.3), confirming that semantic grounding independently improves token predictability.
FTP and SA target different failure modes, since FTP emphasizes local predictability and SA emphasizes semantic invariance.
In practice, the two objectives achieve similar gains independently, and their combination does not yield additive improvement.

\paragraph{Reconstruction improvement is a byproduct.}
Our ablation reveals that the 5.0\% speech Mel improvement over AUV comes from the training procedure (GAN, multi-scale losses, low-LR fine-tuning), not from FTP or SA.
All ablation variants achieve identical reconstruction.
This means the learnability gains (35$\times$ perplexity, +12.1 SALMon) come entirely from LLM-facing objectives, while the reconstruction gains come entirely from the training procedure.
The two effects are independent and additive.

\paragraph{Domain effects.}
The SALMon improvements are speech-specific because SA relies on speech-text correspondence.
For non-speech audio, SA becomes ill-posed or indirect.
This gap motivates future work on alternative supervision for music and environmental sound.

\paragraph{Computation.}
Training requires an extra frozen-LLM forward pass, a GAN discriminator update, and auxiliary prediction heads.
The total training budget is 25k steps.
Inference is unchanged, since the deployed codec keeps the same encoder-decoder structure and discards the auxiliary heads.

\section{Conclusion}

We proposed \ours, which trains a neural audio codec with objectives that reflect downstream autoregressive modeling.
\ours adds future token prediction and semantic alignment through a differentiable bridge, without modifying the codec or LLM architectures.
Experiments show that \ours improves both reconstruction fidelity (5.0\% speech Mel) and token learnability (SALMon +12.1 points, perplexity 35$\times$ lower).
Ablation confirms that these gains are additive: reconstruction improves from the training procedure, while learnability improves from FTP and SA.
These results indicate that tokenizer predictability, not reconstruction quality, is the key bottleneck for spoken language modeling.

\clearpage
\section*{Limitations}

\paragraph{Speech-centric supervision.}
SA relies on speech-text correspondence, so it requires paired transcripts during training.
This assumption holds for read-speech corpora such as LibriSpeech, but it is weaker for untranscribed audio or non-speech domains.
Our cross-domain results reflect this limitation and motivate alternative alignment signals beyond text.

\paragraph{Frozen LLM backbone.}
We freeze the LLM backbone to preserve its text competence and to isolate the effect of tokenizer training.
Jointly adapting the LLM could further improve speech modeling, but it may introduce regressions on text tasks and complicate attribution.

\paragraph{Evaluation coverage.}
Our primary evaluations emphasize read speech.
Conversational settings contain disfluencies, overlap, and rapid speaker turns, which may stress different token properties.
Future work should validate \ours under conversational corpora and multi-speaker conditions.

\paragraph{Training overhead.}
The auxiliary Medusa heads increase training-time memory and compute, even though they are discarded for inference.
This cost may limit scaling to larger backbones or larger batches.

\paragraph{Learnability ablation.}
Our ablation study confirms that reconstruction is invariant to the choice of LLM-facing objectives and prediction horizon $K$.
However, we do not ablate the learnability impact (SALMon, perplexity) of each component separately.
Future work should measure how FTP-only, SA-only, and different $K$ values affect downstream speech coherence.

\section*{Acknowledgement}   
This work was supported by the Ministry of Education (MOE) of Taiwan under the project Taiwan Centers of Excellence in Artificial Intelligence, through the NTU Artificial Intelligence Center of Research Excellence (NTU AI-CoRE).
We thank the National Center for High-performance Computing (NCHC) of the National Applied Research Laboratories (NARLabs) in Taiwan for providing computational and storage resources.

\bibliography{custom}
\appendix
\section{Implementation Details}
\label{app:impl}

This appendix reports the full implementation details of \ours.
This appendix specifies model configurations, training hyperparameters, and stability settings.
This appendix also clarifies how we compute and report reconstruction metrics across domains.

\subsection{Model Configurations}

\paragraph{Codec.}
We use the AUV~\cite{chen2025auv} codec at a 50~Hz token rate.
The codec uses a vocabulary size of 20{,}480.
The codec latent dimension is 256.
The codec hop length is 320.
We fine-tune both the codec encoder and decoder at a low learning rate ($5 \times 10^{-6}$) to allow gradual adaptation while preserving reconstruction quality.

\paragraph{LLM backbone.}
We use Qwen3-4B-Instruct~\cite{yang2025qwen3} as the frozen LLM backbone.
The LLM has 32 transformer layers.
The LLM hidden dimension is 2{,}560.
We extend the LLM vocabulary with 20{,}480 audio tokens via \texttt{resize\_token\_embeddings}.
We train only the audio-token embeddings.
We freeze all other LLM parameters.

\paragraph{Gumbel bridge.}
We implement the Gumbel bridge as a single linear projection from codec latents to audio-token logits.
We use $\texttt{nn.Linear}(256, 20480)$.
We anneal the Gumbel-Softmax temperature from 1.0 to 0.3 over 20k steps.
We use a cosine schedule for annealing.

\paragraph{Medusa heads (FTP).}
We use $K=5$ Medusa-style prediction heads.
Each head is a bias-free linear layer: $\texttt{nn.Linear}(2560, 20480, \texttt{bias=False})$.
We initialize each head from the frozen LLM output projection restricted to audio token indices.
We implement this initialization as \texttt{lm\_head.weight[audio\_ids]}.

\subsection{Training Configuration}
Table~\ref{tab:hyperparams} lists the main training hyperparameters.
We use 4-second audio segments.
We use gradient accumulation to reach an effective batch size of 10.
We train for 25k optimization steps.
We clip the gradient norm to 15.0.
We use SGD with momentum 0.9 for the codec encoder and decoder, and AdamW for the audio token embeddings and Medusa heads.

\begin{table}[h]
\centering
\small
\begin{adjustbox}{width=\columnwidth}
\begin{tabular}{lc}
\toprule
\textbf{Hyperparameter} & \textbf{Value} \\
\midrule
Batch size & 1 \\
Gradient accumulation & 10 \\
Effective batch size & 10 \\
Segment length & 4 seconds \\
\midrule
LR (encoder) & $5 \times 10^{-6}$ \\
LR (decoder) & $5 \times 10^{-6}$ \\
LR (embeddings + heads) & $1 \times 10^{-4}$ \\
Codec optimizer & SGD (momentum$=$0.9, wd$=$1e-4) \\
Embed optimizer & AdamW ($\beta_1{=}0.9$, $\beta_2{=}0.99$) \\
Gradient clip & 15.0 \\
Total steps & 25k \\
Warmup & 2k steps \\
\midrule
$\lambda_{\text{mel}}$ & 1.5 \\
$\lambda_{\text{ms-mel}}$ & 0.5 \\
$\lambda_{\text{mr-stft}}$ & 0.5 \\
$\lambda_{\text{cstft}}$ & 0.8 (phase weight 0.5) \\
$\lambda_{\text{FTP}}$ & 0.2 (after ramp) \\
$\lambda_{\text{cos}}$ & 0.1 (after ramp) \\
$\lambda_{\text{ctr}}$ & 0.05 (after ramp) \\
\midrule
D-only warmup & 0--10k steps \\
FTP schedule & delay 10k, warmup 2k \\
SA schedule & delay 12k, warmup 2k \\
\bottomrule
\end{tabular}
\end{adjustbox}
\caption{Training hyperparameters.}
\label{tab:hyperparams}
\end{table}

\subsection{Semantic Alignment Details}
Table~\ref{tab:sa_hparams} lists the semantic alignment hyperparameters.
We align a mid-to-high layer range to target semantic representations.
We use a memory bank to provide negatives for the contrastive loss.
We use a fixed logit scale and label smoothing.

\begin{table}[h]
\centering
\small
\begin{tabular}{lc}
\toprule
\textbf{Parameter} & \textbf{Value} \\
\midrule
Layer range & $[L/3, 0.8L] = [10, 25]$ \\
Layer weights & Uniform \\
Memory bank size & 512 \\
Logit scale $\alpha$ & 5.0 \\
Label smoothing $\epsilon$ & 0.1 \\
\bottomrule
\end{tabular}
\caption{Semantic alignment hyperparameters.}
\label{tab:sa_hparams}
\end{table}

\subsection{Training Phases}
Training proceeds in three phases over 25k steps.

\paragraph{Phase 1: D-only warmup (steps 0--10k).}
Only the GAN discriminators are updated.
The codec encoder and decoder remain in train mode (so VQ EMA statistics continue tracking), but their optimizer steps are skipped.
Audio token embeddings and Medusa heads are updated throughout all phases.

\paragraph{Phase 2: Full training + FTP (steps 10k--12k).}
The codec encoder and decoder optimizers activate.
FTP loss ramps from zero to full weight over 2k steps.
SA remains inactive.

\paragraph{Phase 3: Full training + FTP + SA (steps 12k--25k).}
SA cosine and contrastive losses ramp from zero to full weight over 2k steps.
All objectives are active for the remaining 11k steps.

This staggered schedule ensures the codec first learns basic reconstruction (Phase 1), then adapts to token predictability (Phase 2), and finally incorporates semantic grounding (Phase 3).

\subsection{Reconstruction Losses}
We train the codec with multiple reconstruction losses.
We use a log mel-spectrogram $\ell_1$ loss with 100 mel bins and hop size 256, with per-sample RMS normalization applied to both input and reconstruction before computing the loss.
We use multi-scale mel losses at FFT sizes 512, 1024, and 2048 with 80 mel bins.
We use a multi-resolution STFT loss combining spectral convergence and log magnitude error.
We use a complex STFT loss that additionally measures phase distance via unit complex vector difference (phase weight 0.5), computed at FFT sizes 512, 1024, and 2048.
We omit STOI and PESQ from the training loss.
We report STOI and PESQ only for evaluation.

\paragraph{Phase jitter.}
We apply random phase jitter ($\pm$24 samples) exclusively to the LM branch input.
This augments the LM pathway with slight temporal perturbations while keeping the reconstruction pathway unaffected, encouraging the encoder to produce tokens that are stable under small phase shifts.

\subsection{GAN Training}
We enable adversarial training with a discriminator-only warmup phase.
The discriminator trains alone for the first 10k steps, allowing it to develop meaningful feature maps before providing gradients to the generator.
We use a multi-period discriminator (MPD) with periods $\{2, 3, 5, 7, 11\}$.
We use a multi-scale discriminator (MSD) with three scales and AvgPool downsampling.
We use hinge loss with feature matching (initial weight 1.5, decaying to 1.0).
We pause GAN updates if the feature-matching loss exceeds 99\% of the total loss, for 500 steps when this condition holds.
We apply R1 regularization with $\gamma=2.0$ every 16 steps.
GAN discriminators run in FP32 precision for stability.

\subsection{Numerical Stability}
We apply several stability mechanisms.
We skip parameter updates when the loss is non-finite.
We clamp logits to $[-80, 80]$ before softmax.
We clip audio samples to $[-1.2, 1.2]$.
We apply code noise by randomly replacing 1.5\% of tokens as regularization for the LM branch.
We set all BatchNorm layers in the codec to eval mode throughout training to prevent batch statistics from drifting.

\section{Additional Results}
\label{app:results}

\subsection{Metric Applicability and Reporting Rationale}
\label{app:metrics}

This section clarifies which reconstruction metrics are meaningful for each domain in Codec-SUPERB-tiny.
Each metric assumes specific signal properties.
We report a metric only when its assumptions hold.
This protocol avoids misleading comparisons across domains.

\paragraph{Mel distance (Mel).}
Mel distance measures the $\ell_1$ error between log mel-spectrograms.
Mel distance is defined for all audio signals.
Mel distance reflects coarse spectral fidelity.
We report Mel for Speech, Music, and Environmental Audio.

\paragraph{STFT distance (STFT).}
STFT distance measures spectral reconstruction error in the short-time Fourier domain.
We use a multi-resolution STFT variant with spectral convergence and magnitude error.
STFT distance is defined for all audio signals.
STFT distance captures finer spectral structure than mel features.
We report STFT for Speech, Music, and Environmental Audio.

\paragraph{Perceptual Evaluation of Speech Quality (PESQ).}
PESQ is calibrated for human speech perception.
PESQ assumes a speech-like signal.
PESQ is therefore most meaningful for Speech and singing voice.
We report PESQ for Speech and Music.
We report PESQ for Environmental Audio for completeness.
We interpret PESQ on Environmental Audio cautiously.

\paragraph{Short-Time Objective Intelligibility (STOI).}
STOI measures speech intelligibility using temporal envelopes in short-time bands.
STOI assumes linguistic content.
STOI is meaningful for Speech and singing voice.
STOI is not meaningful for Environmental Audio.
We report STOI for Speech and Music.
We do not report STOI for Environmental Audio.

\paragraph{Fundamental frequency correlation (F0 Corr).}
F0 correlation measures pitch preservation by correlating estimated fundamental frequency trajectories.
F0 correlation assumes a stable harmonic structure with a well-defined fundamental frequency.
Many environmental sounds are non-harmonic.
Many speech segments contain long unvoiced regions.
F0 correlation is therefore most meaningful for Music.
We report F0 Corr for Music only.

\paragraph{Summary of reported metrics.}
We use the following metric sets:
\begin{itemize}[leftmargin=1.2em, itemsep=0em, topsep=0.2em]
\item Speech: Mel, STFT, PESQ, STOI.
\item Music: Mel, STFT, PESQ, STOI, F0 Corr.
\item Environmental Audio: Mel, STFT, PESQ.
\end{itemize}

\subsection{SALMon Emotion Categories}
\label{app:salmon_emotion}

Table~\ref{tab:salmon_emotion} reports the emotion-related SALMon categories omitted from the main text.

\begin{table}[h]
\centering
\small
\begin{tabular}{lcc}
\toprule
\textbf{Model} & \textbf{Sent-Align} & \textbf{Sent-Cons} \\
\midrule
WavTok-L & 48.0 & 48.0 \\
BigCodec & 53.0 & 48.5 \\
UniCodec & 52.5 & 54.5 \\
AUV & 47.5 & 56.0 \\
\midrule
\textbf{\ours} & 49.5 & \textbf{64.0} \\
\bottomrule
\end{tabular}
\caption{\textbf{SALMon emotion categories.} Sentiment consistency shows a clear gain (+8.0 over AUV). Sentiment alignment is comparable to baselines (49.5\%).}
\label{tab:salmon_emotion}
\end{table}

\subsection{Perplexity Comparison}
\label{app:perplexity}

Table~\ref{tab:ppl_detail} reports token-level perplexity for all codecs after 3 epochs of LM training on LibriSpeech train-clean-100.

\begin{table}[h]
\centering
\small
\begin{tabular}{lcc}
\toprule
\textbf{Model} & \textbf{Eval Loss} & \textbf{Perplexity} \\
\midrule
WavTok-L & 11.91 & 148{,}122 \\
UniCodec & 11.92 & 150{,}197 \\
BigCodec & 11.96 & 156{,}448 \\
AUV & 11.98 & 159{,}768 \\
\midrule
\textbf{\ours} & \textbf{8.44} & \textbf{4{,}617} \\
\bottomrule
\end{tabular}
\caption{\textbf{Token-level perplexity.} All baselines cluster in the 148K--160K range. \ours achieves 4{,}617, a 35$\times$ reduction over AUV.}
\label{tab:ppl_detail}
\end{table}

\subsection{Domain-Specific Detailed Results}
\label{app:domain}

This section reports full reconstruction results for each domain.
Table~\ref{tab:app_overall} averages metrics over all domains.
Tables~\ref{tab:app_speech}, \ref{tab:app_music}, and \ref{tab:app_audio} report domain-specific results.

\begin{table}[h]
\centering
\footnotesize
\setlength{\tabcolsep}{2pt}
\begin{tabular}{lccccc}
\toprule
\textbf{Model} & \textbf{Mel}$\downarrow$ & \textbf{STFT}$\downarrow$ & \textbf{PESQ}$\uparrow$ & \textbf{STOI}$\uparrow$ & \textbf{F0}$\uparrow$ \\
\midrule
UniCodec & \textbf{1.121} & \textbf{2.534} & 1.810 & 0.651 & 0.620 \\
WavTok-M & 1.231 & 2.632 & 1.600 & 0.602 & 0.587 \\
WavTok-S & 1.405 & 2.864 & 1.353 & 0.541 & 0.535 \\
WavTok-L & 1.423 & 2.893 & 1.455 & 0.549 & 0.585 \\
BigCodec & 1.478 & 3.230 & 1.840 & \underline{0.660} & \underline{0.635} \\
\midrule
AUV (base) & \underline{1.246} & 2.871 & \textbf{1.952} & \textbf{0.660} & \textbf{0.650} \\
\textbf{\ours} & 1.229 & \underline{2.843} & \underline{1.946} & \underline{0.664} & 0.630 \\
\bottomrule
\end{tabular}
\caption{\textbf{Overall reconstruction quality} (6,000 samples across all domains). All models operate at 50 tokens/s. UniCodec achieves best spectral fidelity (Mel, STFT). \ours improves over AUV on spectral metrics while remaining competitive on perceptual metrics.}
\label{tab:app_overall}
\end{table}

\begin{table}[h]
\centering
\footnotesize
\setlength{\tabcolsep}{2.5pt}
\begin{tabular}{lcccc}
\toprule
\textbf{Model} & \textbf{Mel}$\downarrow$ & \textbf{STFT}$\downarrow$ & \textbf{PESQ}$\uparrow$ & \textbf{STOI}$\uparrow$ \\
\midrule
BigCodec & 0.810 & 1.718 & \textbf{2.208} & \textbf{0.877} \\
UniCodec & 0.830 & 1.824 & 2.022 & 0.851 \\
WavTok-M & 0.904 & 1.846 & 1.843 & 0.823 \\
WavTok-S & 1.096 & 2.174 & 1.437 & 0.761 \\
WavTok-L & 1.133 & 2.006 & 1.547 & 0.765 \\
\midrule
AUV (base) & \underline{0.762} & \underline{1.648} & 2.094 & 0.850 \\
\textbf{\ours} & \textbf{0.724} & \textbf{1.599} & \underline{2.102} & \underline{0.859} \\
\bottomrule
\end{tabular}
\caption{\textbf{Speech domain results} (2,000 samples from 10 datasets). We omit F0 correlation because pitch accuracy is less central for speech than intelligibility. \ours achieves best spectral fidelity (Mel 0.724, STFT 1.599), improving 5.0\% over AUV. BigCodec leads on perceptual metrics (PESQ 2.208, STOI 0.877).}
\label{tab:app_speech}
\end{table}

\begin{table}[h]
\centering
\footnotesize
\setlength{\tabcolsep}{2pt}
\begin{tabular}{lccccc}
\toprule
\textbf{Model} & \textbf{Mel}$\downarrow$ & \textbf{STFT}$\downarrow$ & \textbf{PESQ}$\uparrow$ & \textbf{STOI}$\uparrow$ & \textbf{F0}$\uparrow$ \\
\midrule
UniCodec & \underline{1.196} & \textbf{2.441} & 1.859 & 0.588 & \underline{0.760} \\
WavTok-M & 1.407 & 2.683 & 1.579 & 0.528 & 0.683 \\
BigCodec & 1.522 & 3.108 & 1.922 & \underline{0.606} & 0.728 \\
WavTok-L & 1.558 & 2.896 & 1.484 & 0.478 & 0.722 \\
WavTok-S & 1.631 & 2.976 & 1.350 & 0.475 & 0.629 \\
\midrule
AUV (base) & \underline{1.129} & 2.557 & \textbf{2.195} & \underline{0.609} & \textbf{0.785} \\
\textbf{\ours} & \textbf{1.126} & \underline{2.541} & \underline{2.175} & \textbf{0.614} & \underline{0.767} \\
\bottomrule
\end{tabular}
\caption{\textbf{Music domain results} (2,000 samples from 6 datasets). AUV leads on PESQ (2.195) and F0 (0.785). \ours slightly improves Mel (1.126 vs.\ 1.129) and STOI (0.614 vs.\ 0.609). UniCodec achieves the lowest STFT distance (2.441).}
\label{tab:app_music}
\end{table}

\begin{table}[h]
\centering
\footnotesize
\setlength{\tabcolsep}{3pt}
\begin{tabular}{lccc}
\toprule
\textbf{Model} & \textbf{Mel}$\downarrow$ & \textbf{STFT}$\downarrow$ & \textbf{PESQ}$\uparrow$ \\
\midrule
UniCodec & \textbf{1.337} & \textbf{3.336} & 1.548 \\
WavTok-M & \underline{1.382} & \underline{3.367} & 1.380 \\
WavTok-S & 1.487 & 3.442 & 1.273 \\
WavTok-L & 1.579 & 3.777 & 1.334 \\
BigCodec & 2.101 & 4.864 & 1.391 \\
\midrule
AUV (base) & 1.847 & 4.407 & \textbf{1.567} \\
\textbf{\ours} & 1.836 & 4.389 & \underline{1.562} \\
\bottomrule
\end{tabular}
\caption{\textbf{Environmental audio results} (2,000 samples from 4 datasets). We omit STOI and F0 correlation because intelligibility and pitch are undefined for environmental sounds. \ours slightly improves over AUV on Mel (1.836 vs.\ 1.847). UniCodec achieves the best spectral metrics.}
\label{tab:app_audio}
\end{table}

\subsection{Ablation Study: Cross-Domain Results}
The main text reports speech-only ablations (Table~\ref{tab:ablation_recon}).
This section reports cross-domain ablations for completeness.
Table~\ref{tab:app_ablation} reports Mel distance across Speech, Music, and Environmental Audio.

\begin{table}[h]
\centering
\footnotesize
\begin{adjustbox}{width=0.5\textwidth}
\begin{tabular}{lccc}
\toprule
\textbf{Variant} & \textbf{Speech Mel}$\downarrow$ & \textbf{Music Mel}$\downarrow$ & \textbf{Audio Mel}$\downarrow$ \\
\midrule
AUV (original) & 0.762 & 1.129 & 1.847 \\
\midrule
FTP only & 0.725 & 1.126 & 1.836 \\
SA only & 0.723 & 1.126 & 1.837 \\
\midrule
\ours (FTP + SA) & 0.724 & 1.126 & 1.836 \\
\bottomrule
\end{tabular}
\end{adjustbox}
\caption{\textbf{Cross-domain ablation (Mel distance).} All variants improve $\sim$5\% over the original AUV on Speech, but are indistinguishable from each other ($\leq$0.002). The improvement comes from the shared training procedure.}
\label{tab:app_ablation}
\end{table}

\subsection{Evaluation Datasets}
Table~\ref{tab:datasets} lists the datasets in Codec-SUPERB-tiny~\cite{wu2024codecsuperb}.
We uniformly sample 2{,}000 clips per domain.
This protocol yields 6{,}000 total evaluation samples.

\begin{table}[h]
\centering
\small
\begin{tabular}{ll}
\toprule
\textbf{Dataset} & \textbf{Features} \\
\midrule
\multicolumn{2}{l}{\textit{Speech (10 datasets, 2,000 samples)}} \\
LibriSpeech & diverse speaker, read audiobooks \\
VoxCeleb1 & diverse speaker, celebrities on YouTube \\
Speech Commands v1 & spoken keyword commands \\
QUESST & multi-lingual, low resource language \\
VoxLingua107 Top 10 & multi-lingual, YouTube content \\
Audio SNIPS & spoken commands, crowdsourced \\
IEMOCAP & affective speech \\
CREMA-D & affective speech \\
Libri2Mix & multi-speaker scenarios \\
LibriCount & multi-speaker scenarios \\
\midrule
\multicolumn{2}{l}{\textit{Environmental Audio (4 datasets, 2,000 samples)}} \\
ESC-50 & diverse audio source \\
FSD-50K & diverse audio source \\
Gunshot Triangulation & diverse audio source \\
Vocal Imitations & human imitation of sound \\
\midrule
\multicolumn{2}{l}{\textit{Music (6 datasets, 2,000 samples)}} \\
OpenSinger & singing voice, Chinese song \\
M4Singer & singing voice, Chinese song \\
VocalSet & singing skill \\
NSynth & instrument notes \\
GTZAN Genre & diverse music genre \\
GTZAN Music Speech & instrument note \\
\bottomrule
\end{tabular}
\caption{Datasets in Codec-SUPERB-tiny. We uniformly sample 2,000 clips per domain for balanced evaluation.}
\label{tab:datasets}
\end{table}

\subsection{Model Nomenclature}
We use short names for readability in tables.
We list the corresponding model identifiers below.
\begin{itemize}[leftmargin=1.2em, itemsep=0em, topsep=0.2em]
\item \textbf{WavTok-L}:
\resizebox{0.5\linewidth}{!}{\texttt{wavtokenizer\_24k\_large\_600\_4096}}
\item \textbf{WavTok-M}:
\resizebox{0.5\linewidth}{!}{\texttt{wavtokenizer\_24k\_medium\_600\_4096}}
\item \textbf{WavTok-S}:
\resizebox{0.5\linewidth}{!}{\texttt{wavtokenizer\_24k\_small\_600\_4096}}
\end{itemize}

\subsection{Qualitative Examples}
We provide audio samples in the supplementary material.
The supplementary material includes:
\begin{itemize}[leftmargin=1.2em, itemsep=0em, topsep=0.2em]
\item Reconstruction comparisons across codecs.
\item Generation samples for baseline versus \ours tokenizers.
\item Robustness examples under noise and perturbations.
\end{itemize}

\end{document}